%% file: Petersson2015.tex
\pdfoutput=1
\documentclass[10pt]{iopart}
\usepackage[utf8]{inputenc}
\usepackage{iopams}
\usepackage{defs}
\usepackage[square,sort&compress,numbers]{natbib}

\usepackage[font=small, labelfont=bf]{caption}
\usepackage{float}
\input{tikzcfg}
\newcommand{\eqref}[1]{(\ref{#1})}

\begin{document}
\title{Phase metrology with multi-cycle two-colour pulses}

\author{C~L~M~Petersson$^{1,2}$, S~Carlstr\"{o}m$^1$, K~J~Schafer$^3$ and
  J~Mauritsson$^1$}
\address{$^1$ Department of Physics, Lund University, P. O. Box 118,
  S-22100 Lund, Sweden}
\address{$^2$ Departamento de Química, Modulo 13, Universidad Autónoma de
  Madrid, 28049, Madrid, Spain}
\address{$^3$ Department of Physics and Astronomy, Louisiana State
  University, Baton Rouge, LA 70803}
\ead{leon.petersson@uam.es}

\begin{abstract}
  Strong-field phenomena driven by an intense infrared (IR) laser
  depend on during what part of the field cycle they are initiated. By
  changing the sub-cycle character of the laser electric field it is
  possible to control such phenomena. For long pulses, sub-cycle
  shaping of the field can be done by adding a relatively weak, second
  harmonic of the driving field to the pulse. Through constructive and
  destructive interference, the combination of strong and weak fields
  can be used to change the probability of a strong-field process
  being initiated at any given part of the cycle. In order to control
  sub-cycle phenomena with optimal accuracy, it is necessary to know
  the phase difference of the strong and the weak fields precisely. If
  the weaker field is an even harmonic of the driving field, electrons
  ionized by the field will be asymmetrically distributed between the
  positive and negative directions of the combined fields. Information
  about the asymmetry can yield information about the phase
  difference. A technique to measure asymmetry for few-cycle pulses,
  called Stereo-ATI (Above Threshold Ionization), has been developed
  by [Paulus G~G, \etal 2003 {\em Phys. Rev. Lett.\/} {\bf 91}].  This
  paper outlines an extension of this method to measure the phase
  difference between a strong IR and its second harmonic.
\end{abstract}


\submitto{\jpb} \maketitle

\section{Introduction}
\label{sec:Introduction}

Strong field processes such as high-order harmonic generation (HHG)
and above threshold ionization (ATI) depend on the sub-cycle structure
of the strong infrared (IR) field driving the process. By tailoring
the sub-cycle structure of the field, one can control the
processes. This can be done either by using very short laser pulses
\citep{Perry1993,Eichmann1995,Dudovich2006} or by mixing pulses with
different colours
\citep{Mauritsson2009JoPBAMaOP,Nguyen2004PRA,brizuela2013SR}. HHG with
few-cycle laser pulses has enabled the generation of isolated
attosecond pulses \citep{Hentschel2001} and in this case the process
is controlled by changing the so-called carrier--envelope phase (CEP)
\citep{Reichert1999,Udem2002,Goulielmakis2004,Brabec2000,Jones2000}:
\begin{equation}
  \label{eq:cep}
  \vec{E}(t) = \vec{E}_0f(t)\sin(\omega t+\phi_{\mathrm{CEP}}),
\end{equation}
where \(f(t)\)
describes the envelope of the pulse with respect to which
\(\phi_{\mathrm{CEP}}\) is measured.

\begin{figure}[H]
  \centering
  \includegraphics{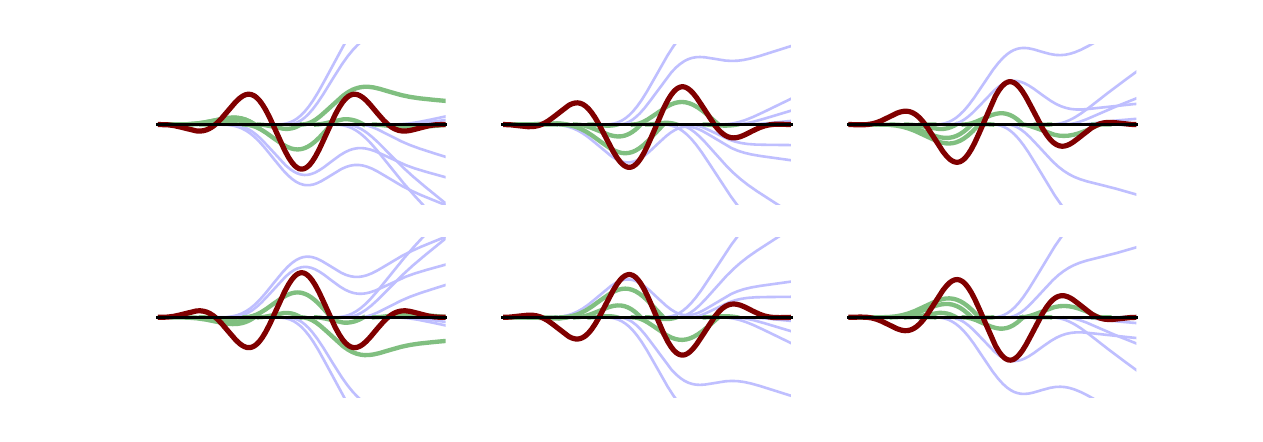}
  \caption{An illustration of the classical paths of electrons
    ionized by a few-cycle pulse, for six different CEPs. The green
    paths lead the electrons back to the atom, opening for the
    possibility of rescattering, whereas the blue paths guarantee
    direct ionization. The CEP of the driving field, shown in red,
    is changed by increments of $\frac{1}{3}\pi$ between each
    figure, giving phase difference of $\pi$ between the two
    pictures in each column. Note the asymmetry between the rows.}
  \label{fig:few-cycle-paths}
\end{figure}

The CEP relates the phase of the driving frequency to the envelope of
the pulse and changing it may lead to the generation of one or two
attosecond pulses if the duration is sufficiently short
\citep{Baltuska2003}. The rapid change of amplitude of a short pulse
broadens the pulse frequency distribution and breaks the symmetrical
distribution of electron paths between the positive and negative
directions of the field \citep{Paulus2001N,Paulus2003}. How this
depends on the CEP is illustrated in figure \ref{fig:few-cycle-paths}.

\begin{figure}
  \centering
  \tikzsetnextfilename{figure2}
  \begin{tikzpicture}
    \foreach \r/\c/\offset/\placement/\angle in {0/0/0/0/0, 0/1/339/360/$\frac{\pi}{2}$, 1/0/180/360/$\pi$, 1/1/159/360/$\frac{3\pi}{2}$}{
      \def\minx {\offset}
      \def\maxx {3*360+\offset}
      \begin{axis}[
        clip bounding box=upper bound,
        at={ ($ (5cm,0) + (1.05*\c*\textwidth/2,-\r*\textwidth/6.5) $) },
        height=\textwidth/4,
        width= \textwidth*0.5/2,
        samples=200,
        axis line style=thick,
        ticks=none,
        xmin=-1,
        xmax=360+1,
        axis x line*=center,
        ymin=-1.55,
        ymax=1.55,
        hide y axis,
        ]
        \addplot[red,thick,domain=0:360] {(sin(x)};
        \addplot[blue,thick,domain=0:360] {(sin(2*x + 180*\r + 90*\c)/2};
      \end{axis}
      \begin{axis}[
        clip bounding box=upper bound,
        at={ ($ (5.25cm,0) + (1.05*\c*\textwidth/2,-\r*\textwidth/6.5) + (\textwidth*0.25/2,0) $) },
        height=\textwidth/4,
        width= \textwidth*0.76/2,
        axis line style=thick,
        samples=200,
        ticks=none,
        xmin=\minx-3,
        xmax=\maxx+3,
        axis x line*=center,
        ymin=-1.55,
        ymax=1.55,
        hide y axis,
        ]
        \addplot[blue!20!purple!50!white, thick, domain=\minx:\maxx] {(sin(x) + sin(2*x + 180*\r + 90*\c)/2};
        \addplot[blue!20!purple!50!black, thick, domain=\placement:\placement+360] {(sin(x) + sin(2*x + 180*\r + 90*\c)/2};
      \end{axis}
      \node at ($ (4.5cm,0.84cm) + (1.05*\c*\textwidth/2,-\r*\textwidth/6.5) $) {\small$\phi=$\angle};
    }
  \end{tikzpicture}
  \caption{A constant-amplitude field is shown together with its
    weaker, second harmonic for four different values of $\phi$. For
    each value of $\phi$ the first and the second harmonics are
    shown, in red and blue respectively, to the left, and their sum
    in purple to the right. To the right an interval corresponding
    to that in the left figure is shown in a darker colour.}
  \label{fig:two-colour-asymmetry}
\end{figure}
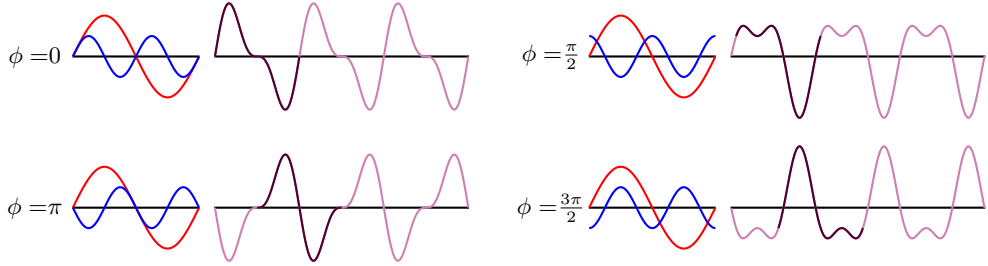

For two-coloured, multi-cycle fields with commensurate frequencies,
the total electric is given by
\begin{equation}
  \label{eq:field-amplitude}
  \vec{E}(t)=\vec{E}_1\sin(\omega t) + \vec{E}_2\sin(n\omega t+\phi),
\end{equation}
where \(\vec{E}_1,\vec{E}_2\)
are the amplitudes of the fundamental frequency and its \(n\)th
harmonic (in this letter, \(n=2\)),
respectively, and \(\phi\)
is the relative phase between the two fields. The CEP is neglected in
\eqref{eq:field-amplitude}, since it has a negligible effect on the
symmetry for multi-cycle pulses. Changing the relative phase of the
two-coloured field may for instance result in one or two pulses per
cycle \citep{Mauritsson2006} (see figure \ref{fig:two-colour-asymmetry}).  One
would assume that maximizing the asymmetry would also maximize the
harmonic yield, since the harmonic yield scales with the field
maximum, which is maximized at biggest asymmetry. However, SFA
calculations showed \citep{Mauritsson2006} that there is a phase
offset between maximum asymmetry and maximum harmonic yield and the
highest harmonic cutoff.  To truly understand and measure the impact
of the sub-cycle structure, and enable the comparison between theory
and experiment, the relative phase has to be measured independently
from the process being studied and it has to be measured ``on
target'', where the harmonics are being generated.

For few-cycle pulses the phase is measured using a method known as
Stereo-ATI \citep{Wittmann2009NP, Rathje2012JoPBAMaOP}. In Stereo-ATI
the direction of the ionized electrons is measured and related to the
CEP. In this letter we show that the Stereo-ATI technique can be used
to measure the relative phase of a two-coloured field with
commensurate frequencies and also the relative strength of the two
fields, since varying \(\phi\)
has an impact on the field structure very similar to that of the CEP
on short pulses \citep{Mauritsson2009JoPBAMaOP}.

\subsection{Multi-cycle pulses}
\begin{figure}[t]
  \centering
  \tikzsetnextfilename{figure3}
  \begin{tikzpicture}
    \begin{scope}[xshift=-4cm]
      \node at (0,1.5) {\input{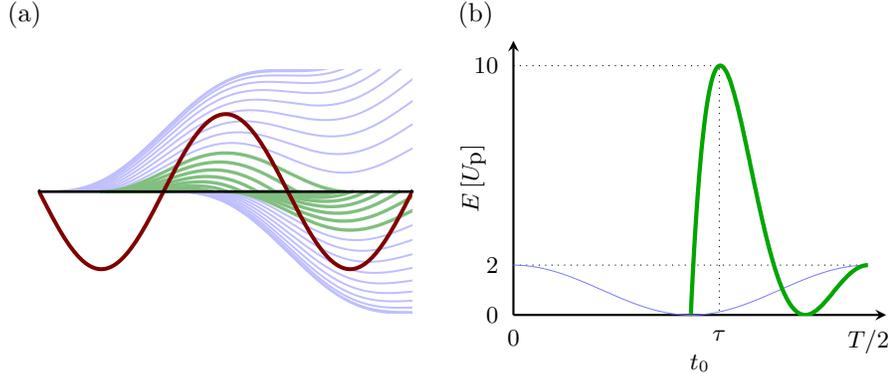}};
      \node at (-2.5,4) {(a)};
    \end{scope}
    \begin{scope}
      \node at (-0.5,4) {(b)};
      \begin{axis}[
        width=\textwidth/2,
        height=\textwidth/2.5,
        axis line style=thick,
        xmin=0,
        xmax=2100,
        axis x line=center,
        xtick={1,1160,2000},
        xticklabels={0,$\tau$,$T/2$},
        xtick style={draw=none},
        xlabel={$t_0$},
        xlabel style={at=(current axis.south), below=4mm},
        ymin=0, ymax=11, ytick={0,2,10}, ytick style={draw=none}, axis y
        line=left, ylabel={\small$E\,$[$\Up$]}, ylabel shift = -0.4cm, ]
        \addplot[dotted,black] coordinates{(0.01,2) (2000,2)};
        \addplot[dotted,black] coordinates{(0,10) (1160,10)};
        \addplot[dotted,black] coordinates{(1160,-0.079) (1160,10)};
        \addplot[no marks,blue!50!white] table[x index=0,y index=1] {di_figure3b.txt};
        \addplot[no marks,ultra thick,green!65!black] table[x index=0,y index=1] {rs_figure3b.txt};
      \end{axis}
    \end{scope}
  \end{tikzpicture}
  \caption{In (a), the classical paths of electrons ionized during one
    cycle of a laser field, shown in red, is displayed. Analogously to
    figure \ref{fig:few-cycle-paths}, the green paths lead back to the core,
    giving a possibility of rescattering, whereas the blue ensure
    direct ionization. In (b), the energy $E$
    of the directly ionized electrons is shown in blue, and the
    maximum classical energy of the rescattered electrons is shown in
    green, both as a function of the ionization time, $t_0$,
    over one half cycle of the field. Classically, electrons can only
    gain energies of $10\Up$ if they are ionized at time $\tau$.}
  \label{fig:classical-paths-final-energy}
\end{figure}

For multi-cycle pulses, the asymmetry due to variations in amplitude
is very small from half-cycle to half-cycle. By approximating the
amplitude as constant, the path of a directly ionized electron,
leaving the atom at time $t_0$,
can be seen as depending on the instant of ionization. This also
decides its final energy, which is the same as the energy of an
electron leaving the atom one half-cycle later, but in the opposite
direction so that they do not overlap. The classical paths of
electrons ionized by a strong one-coloured field are shown in figure
\ref{fig:classical-paths-final-energy}(a). In this figure, the blue lines correspond to
directly ionized electrons, which may reach a maximum energy of
\(2\Up\), where \(\Up\) is the Ponderomotive energy given by
\begin{equation}
  \Up = \frac{q^2|\vec{E}|^2}{4m\omega^2},
\end{equation}
where $q$
is the elementary charge, $|\vec{E}|$
the field amplitude, $m$
the electron mass and $\omega$
the field frequency.  The green lines, instead, correspond to
electrons that return to the ion core, where they may rescatter. The
maximum energy attainable for the rescattered electrons is reached by
elastic scattering when the velocity is completely reversed
\citep{Paulus1995PRA}. In this way, the electron may reach final
electron energies up to \(10\Up\).
Figure \ref{fig:classical-paths-final-energy}(b) shows the energy of directly ionized
electrons and the maximum classical energy of rescattered electrons as
a function of ionization time.

The acceleration of the free electrons is proportional to the field
strength, and when the second harmonic is much weaker than the driving
field, the electron paths through the field are approximately the same
as for the monochromatic case \citep{Dahlstroem2009PRA,
  Dahlstroem2011JoPBAMaOP}. The non-linear ionization probability,
however, is significantly influenced by the second harmonic
\citep{brizuela2013SR}. As the electron energy depends on the
ionization time, the same principle that was used for short pulses can
be utilized for the two-colour case \citep{Yin1992PRL,
  Nguyen2004PRA}. This paper outlines a method for measuring the phase
difference between the first and second harmonic of a multi-cycle,
two-coloured field, by studying the asymmetry of the ATI spectrum.

\section{Numerical computations}

The calculations were done using a newly developed version of the code
described in \citep{Schafer2009}, designed to run on graphical
processing units. It solves the time-dependent Schr\"{o}dinger
equation in the single active electron approximation and a combined
basis consisting of a radial grid and spherical harmonics.
\begin{figure}[b]
    \center
    \tikzsetnextfilename{figure4}
    \begin{tikzpicture}[
            declare function={
                func(\x)=0 + 
                and(\x>0, \x<=1) * \x + 
                and(\x>1, \x<=9) * 1 +
                and(\x>9,\x<=10) * (10 - \x);
            }
           ]
        \begin{axis}[
                samples=200,
                axis line style=thick,
                height=\textwidth/4.5,
                width=\textwidth,
                axis x line=center,
                x axis line style=-,
                xmin=0,
                xmax=10,
                xtick=\empty,
                hide y axis,
                ymin=-1.1,
                ymax=1.1,
            ]
            \addplot [red!50!white,thick,domain=0:10] {sin(135*x)*func(x)};
            \addplot [blue!50!white,dashed,thick,domain=0:10] {sin(270*x+45)*func(x)/2};
            \addplot [blue!50!white,thick,domain=0:10] {sin(270*x)*func(x)/2};
            \addplot [blue!50!black,thick,domain=0:10] {func(x)/2};
            \addplot [red!50!black,thick,domain=0:10] {func(x)};
        \end{axis}
    \end{tikzpicture}
    \caption{An illustration of how the pulses are simulated. The
      trapezoidal envelopes for the high- and low-frequency pulses are
      shown in dark blue and red, respectively. The relative intensity
      is exaggerated for illustrative purposes. In light red, the
      low-frequency pulse is shown; in light blue, the high-frequency
      pulse is shown as a continuous line for $\phi=0$ and as a dashed
      line for $\phi=\frac{\pi}{4}$.}
    \label{fig:trapezoidal-pulse-schematic}
\end{figure}
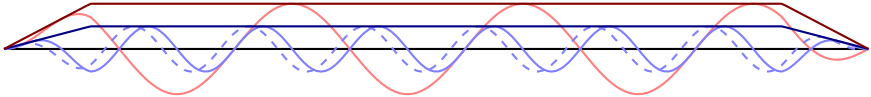

The pulses were modeled using trapezoidal pulse envelopes. This is
advantageous as the asymmetry due to the frequency mixing is present
during a majority of the simulated pulse. To study the asymmetry for
different $\phi$, the CEP of the high-frequency wave was varied
between pulses. Due to the relatively low intensity of the
high-frequency pulse, changes to the asymmetry due to boundary effects
caused by changing CEP are relatively small. The simulated pulse is
illustrated in figure \ref{fig:trapezoidal-pulse-schematic}.

\section{Theory}
\subsection{Asymmetry as a basis for phase metrology}
The force, $\vec{F}$, on an electron in a two-colour laser field can
for non-relativistic velocities be approximated as
\begin{equation}
    \vec{F} = -q\vec{E}_{\omega}\left[\sin\left(\omega t\right) + \sqrt{I_{\textrm{rel}}}\sin\left(2\omega t + \phi\right)\right],
\end{equation}
where $\vec{E}_{\omega}$ is the envelope of the driving field,
$I_{\textrm{rel}}$ the relative intensity of the second harmonic, $q$
the elementary charge, $\omega$ the driving field frequency. The
carrier waves are shown for four different values of $\phi$ in figure
\ref{fig:two-colour-asymmetry}.

\begin{figure}[b]
    \centering
    \tikzsetnextfilename{figure5}
    \begin{tikzpicture}[
            declare function={
                func(\x)=(\x<=1) * (4-\x) + 
                and(\x>1, \x<=3) * (1/4*(\x - 3)^2 + 2) + 
                and(\x>3, \x<=9) * 2 +                  
                and(\x>9,\x<=10) * (2 - 1/4*(\x - 9)^2) + 
                         (\x>10) * (13.5-\x)/2;
            }]
        \begin{axis}[
                samples=200,
                axis line style=thick,
                height=\textwidth/3,
                width=\textwidth,
                axis x line=bottom,
                xmin=0,
                xmax=11,
                xtick={0,1,2,3,4,5,6,7,8,9,10},
                xtick style={draw=none},
                xlabel=${\small E \, [\Up]}$,
                xlabel style={at=(current axis.right of origin), anchor=north},
                axis y line=left,
                ymin=0,
                ymax=5,
                ytick=\empty,
                ylabel=${\small\log(P)}$,
                ylabel shift = -0.1cm,
            ]
            \foreach \E/\P in {1/3, 2/2.25, 3/2, 4/2, 5/2, 6/2, 7/2, 8/2, 9/2, 10/1.75}{
                \addplot[blue,thick] coordinates {(\E,0) (\E,\P)};
            };
            \addplot[red,thick,domain=0:10.75] {func(x)};
        \end{axis}
    \end{tikzpicture}
    \caption{A simplified version of the general characteristics of an ATI spectrum is shown in red, and a division of $\left[0,10\Up\right]$ into ten sections, is shown in blue.}
    \label{fig:schematic-ati}
\end{figure}
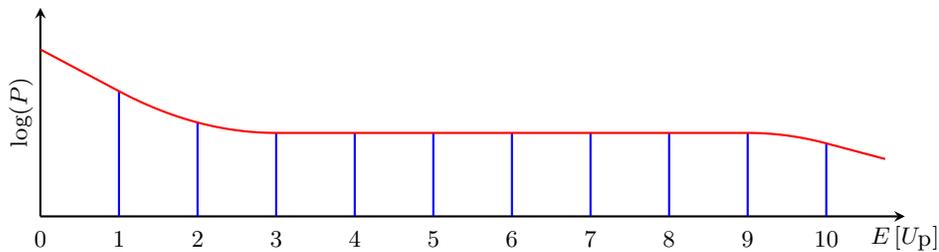

In figure \ref{fig:schematic-ati} a schematic ATI spectrum is shown in
red. In the monochromatic case, there would be an approximate symmetry
between the directions of the field. The addition of the second
harmonic, however, breaks that symmetry. 

The energy of the electrons depend on when during the half-cycle of
the field they are ionized, as shown in the right of figure
\ref{fig:classical-paths-final-energy}. This means that the energy distribution of
electrons in the positive direction of the field, $P_+\left(E\right)$,
will be different from that in the negative direction of the field,
$P_-\left(E\right)$.

If $\sin\left(\omega t\right)$ and $\sin\left(2\omega t + \phi\right)$
constructively interfere at $t=t^\prime$\, they interfere
destructively at $t=t^\prime+\pi$. This is shown in figure
\ref{fig:two-colour-asymmetry}. As a result, the parts of the sub-half cycle when
the driving field is positive, for which there is constructive
interference, are the same parts of the sub-half cycle when the
driving field is negative, for which the interference is destructive,
and vice versa. This will in turn mean that an overrepresentation, due
to constructive interference, of electrons with energy $E$ in one
direction of the field will coincide with an underrepresentation, due
to destructive interference, in the other.

In order to gain information about $\phi$, the asymmetry between
$P_+\left(E\right)$ and $P_-\left(E\right)$ can be
studied. Analogously with \cite{Rathje2012JoPBAMaOP} and
\cite{Wittmann2009NP}, the asymmetry of two energy ranges,
$\varepsilon_\mathrm{l}$ and $\varepsilon_\mathrm{h}$, of the ATI
spectrum will be studied in this paper. The subscripts $\mathrm{l}$
and $\mathrm{h}$ will below be used to differentiate between the low
and the high energy range.

To provide metrics for the respective asymmetries of
$\varepsilon_\mathrm{l}$ and $\varepsilon_\mathrm{h}$,
$A_\mathrm{l} = A\left(\varepsilon_\mathrm{l}\right)$ and
$A_\mathrm{h} = A\left(\varepsilon_\mathrm{h}\right)$ were used, where
\begin{equation}
    A\left(\varepsilon\right) = \frac{\displaystyle\int_{\varepsilon}\mathrm{d}E\left[P_{+}\left(E\right) - P_{-}\left(E\right)\right]}{\displaystyle\int_{\varepsilon}\mathrm{d}E\left[P_{+}\left(E\right) + P_{-}\left(E\right)\right]}
\end{equation}
is the asymmetry over an energy interval $\varepsilon$ of the ATI
spectrum.

\begin{figure}[H]
    \centering
    \tikzsetnextfilename{figure6a_6c}
    \begin{tikzpicture}[
            declare function={
                Al(\x)=0.75*cos(\x-30) -
                0.4*sin(\x-30);
            },
            declare function={
                Ah(\x)=0.55*cos(\x-30) +
                0.4*sin(\x-30);
            },
        ]
        \begin{axis}[ 
                at={ (0*\textwidth/3, 0) },
                width=\textwidth/2.4,
                height=\textwidth/2.4,
                axis line style=thick,
                ticks=none,
                xmin=-1.2,
                xmax=1.2,
                axis x line=center,
                xlabel=$A_\mathrm{l}$,
                xlabel style={at=(current axis.right of origin), anchor=north},
                ymin=-1.2,
                ymax=1.2,
                axis y line=center,
                ylabel=$A_\mathrm{h}$,
                ylabel style={at=(current axis.above origin), anchor=west},
        ]
            \addplot [black,thick, only marks, mark size = 2.5]  coordinates{
                ({Al(0)},{Ah(0)})
                ({Al(60)},{Ah(60)})
                ({Al(120)},{Ah(120)})
                };
            \addplot [blue!50!white,thick, only marks, mark size = 2]  coordinates{
                ({Al(0)},{Ah(0)})
                ({Al(60)},{Ah(60)})
                ({Al(120)},{Ah(120)})
                };
            \node at (axis cs:{Al(0)},{Ah(0)}) [anchor=south west] {\small$0$};
            \node at (axis cs:{Al(60)},{Ah(60)-0.06}) [anchor=south west] {\large$\frac{\pi}{3}$};
            \node at (axis cs:{Al(120)},{Ah(120)}) [anchor=south east] {\large$\frac{2\pi}{3}$};
            \node at (axis cs:-1.075,1.1) [anchor=center] {\small(a)};
        \end{axis}
        \begin{axis}[ 
                at={ (1*\textwidth/3, 0) },
                width=\textwidth/2.4,
                height=\textwidth/2.4,
                axis line style=thick,
                ticks=none,
                xmin=-1.2,
                xmax=1.2,
                axis x line=center,
                xlabel=$A_\mathrm{l}$,
                xlabel style={at=(current axis.right of origin), anchor=north},
                ymin=-1.2,
                ymax=1.2,
                axis y line=center,
                ylabel=$A_\mathrm{h}$,
                ylabel style={at=(current axis.above origin), anchor=west},
        ]
            \addplot [black,thick, only marks, mark size = 2.5]  coordinates{
                ({Al(0)},{Ah(0)})
                ({Al(60)},{Ah(60)})
                ({Al(120)},{Ah(120)})
                ({Al(180)},{Ah(180)})
                ({Al(240)},{Ah(240)})
                ({Al(300)},{Ah(300)})
                };
            \addplot [blue!50!white,thick, only marks, mark size = 2]  coordinates{
                ({Al(0)},{Ah(0)})
                ({Al(60)},{Ah(60)})
                ({Al(120)},{Ah(120)})
                ({Al(180)},{Ah(180)})
                ({Al(240)},{Ah(240)})
                ({Al(300)},{Ah(300)})
                };
            \node at (axis cs:{Al(0)},{Ah(0)}) [anchor=south west] {\small$0$};
            \node at (axis cs:{Al(60)},{Ah(60)-0.06}) [anchor=south west] {\large$\frac{\pi}{3}$};
            \node at (axis cs:{Al(120)},{Ah(120)}) [anchor=south east] {\large$\frac{2\pi}{3}$};
            \node at (axis cs:{Al(180)},{Ah(180)}) [anchor=north east] {\small$\pi$};
            \node at (axis cs:{Al(240)},{Ah(240)+0.06}) [anchor=north east] {\large$\frac{4\pi}{3}$};
            \node at (axis cs:{Al(300)},{Ah(300)}) [anchor=north west] {\large$\frac{5\pi}{3}$};
            \node at (axis cs:-1.075,1.1) [anchor=center] {\small(b)};
        \end{axis}
        \begin{axis}[ 
                at={ (2*\textwidth/3, 0) },
                width=\textwidth/2.4,
                height=\textwidth/2.4,
                ticks=none,
                xmin=-1.2,
                xmax=1.2,
                hide x axis,
%
                ymin=-1.2,
                ymax=1.2,
                hide y axis,
        ]
            \addplot[
                    domain = 0:360,
                    very thick,
                    samples = 200,
                ] ({Al(x)},{Ah(x)});
            \addplot[
                    domain = -24:70,
                    very thick,
                    red!80!white,
                    samples = 50,
                ] ({Al(x)},{Ah(x)});
            \addplot[
                    domain = 0:65,
                    very thick,
                    red!80!white,
                    samples = 50,
                ] ({cos(x)/4},{sin(x)/4});
            \addplot [red!80!white,very thick] coordinates {
                (0,0) 
                ({Al(70)},{Ah(70)})
            };

            \addplot [red!80!white,thick, only marks, mark size = 2.5]  coordinates{
                ({Al(0)},{Ah(0)})
                ({Al(60)},{Ah(60)})
                };
            \addplot [black,thick, only marks, mark size = 2.5]  coordinates{
                ({Al(120)},{Ah(120)})
                ({Al(180)},{Ah(180)})
                ({Al(240)},{Ah(240)})
                ({Al(300)},{Ah(300)})
                };
            \addplot [blue!50!white,thick,only marks,mark size = 2]  coordinates{
                ({Al(0)},{Ah(0)})
                ({Al(60)},{Ah(60)})
                ({Al(120)},{Ah(120)})
                ({Al(180)},{Ah(180)})
                ({Al(240)},{Ah(240)})
                ({Al(300)},{Ah(300)})
                };
            \node at (axis cs:-1.075,1.1) [anchor=center] {\small(c)};
            \node at (axis cs:0.17,0) [red!80!white,anchor=south west] {\small$\theta(\phi)$};
            \node at (axis cs:{Al(70)},{Ah(70)}) [red!80!white,anchor=south] {\small$\phi$};
        \end{axis}
        \begin{axis}[ 
                at={ (2*\textwidth/3, 0) },
                width=\textwidth/2.4,
                height=\textwidth/2.4,
                axis line style=thick,
                ticks=none,
                xmin=-1.2,
                xmax=1.2,
                axis x line=center,
                xlabel=$A_\mathrm{l}$,
                xlabel style={at=(current axis.right of origin), anchor=north},
                ymin=-1.2,
                ymax=1.2,
                axis y line=center,
                ylabel=$A_\mathrm{h}$,
                ylabel style={at=(current axis.above origin), anchor=west},
        ]
        \end{axis}
    \end{tikzpicture}
    \caption{An illustration of the $A_\mathrm{l}$--$A_\mathrm{h}$
      plane. In (a), the representation of a two-colour pulse has been
      given for
      $\phi\in\left\{0,\frac{\pi}{2},\frac{2\pi}{3}\right\}$. In (b),
      an extrapolation of the values in (a) based on symmetry can be
      seen. In (c), the $\phi$-dependent angle $\theta$ is shown.}
    \label{fig:al-ah-plane-schematic}
\end{figure}
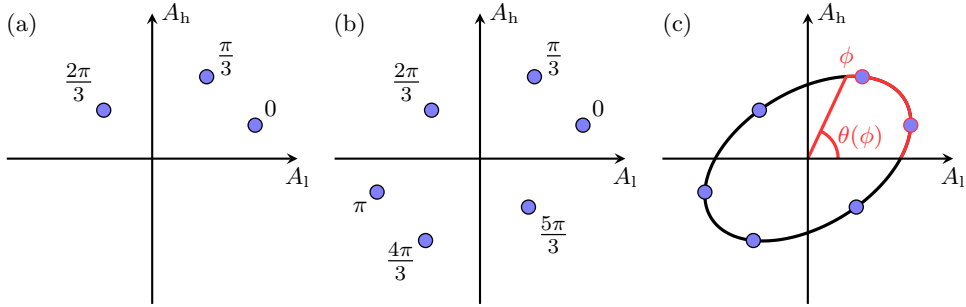

For different values of $\phi$, a two-colour pulse can be represented
in the $A_\mathrm{l}$--$A_\mathrm{h}$ plane. Figure
\ref{fig:al-ah-plane-schematic}(a) illustrates this for a hypothetical wave and
$\phi\in\left\{0,\frac{\pi}{3},\frac{2\pi}{3}\right\}$. As can be seen
in figure \ref{fig:two-colour-asymmetry}, changing the value of $\phi$ by $\pi$
completely inverts the asymmetry coming from the second
harmonic. Because of this, figure \ref{fig:al-ah-plane-schematic}(a) can be
extrapolated to give the values in figure \ref{fig:al-ah-plane-schematic}(b).

Denote the angular coordinate in the $A_\mathrm{l}$--$A_\mathrm{h}$
plane $\theta$. As is shown in figure \ref{fig:al-ah-plane-schematic}(c), there
exists for certain $A_\mathrm{l}$--$A_\mathrm{h}$ representations a
bijective mapping between $\phi$ and $\theta$. By choosing
$\varepsilon_\mathrm{l}$ and $\varepsilon_\mathrm{h}$ that result in
such a mapping, it is possible to gain a measure of $\phi$.

It interesting to note that neither $\phi\rightarrow A_\mathrm{l}$ or
$\phi\rightarrow A_\mathrm{h}$ are injective, which can easily be seen
in figure \ref{fig:al-ah-plane-schematic}(c). As injectivity is a requirement for
inversion, it would not be possible to determine a non-ambiguous
measure of $\phi$ by observing the asymmetry of a single range of the
ATI spectrum, which justifies the previous selection of two energy
ranges.

\subsection{Measurement of the absolute phase difference}
The second harmonic gives rise to constructive and destructive
interference during predetermined parts of each cycle. During
experiments it is important to be certain of which data point in the
$A_\mathrm{l}$--$A_\mathrm{h}$ plane corresponds to which
$\phi$. However, even if the values of $\theta$ in the mapping
\begin{equation}
    \theta\left(\phi\right)\in\left\{\theta\left(\phi_0\right), \theta\left(\phi_0+\frac{2\pi}{N}\right), \cdots , \theta\left(\phi_0-\left(N-1\right)\frac{2\pi}{N}\right) \right\}
\end{equation}
have been ascertained for given $\varepsilon_\mathrm{l}$ and
$\varepsilon_\mathrm{h}$ by changing $\phi$ in increments of
$\frac{2\pi}{N}$, it can be risky to speculate on the value of
$\phi_0$.

One solution to this problem is found in the right hand side of figure
\ref{fig:classical-paths-final-energy}, which shows that there is only one ionization time
per half-cycle, here called $\tau$, for which electrons can
classically obtain energies as high as $10\Up$. Due to quantum
mechanical effects, it is possible for electrons with other ionization
times to obtain equally high energies, but the probability of doing so
is small for ionization times which notably differ from
$\tau$. Because of this, the highest asymmetry near $10\Up$ is
observed when the peak of the second harmonic occur at $\tau$. In
other words, for some small $\delta$,
$A\left(\left[10\Up-\delta,10\Up+\delta\right]\right)$ will be
maximized and positive when the peak of the second harmonic in the
positive field direction occurs at $\tau$. The peak of the second
harmonic occurring at $\tau+\pi$, on the other hand, maximizes
$-A\left(\left[10\Up-\delta,10\Up+\delta\right]\right)$ .

\subsection{Selection of $\varepsilon_\mathrm{l}$ and $\varepsilon_\mathrm{h}$}

Because the asymmetries of $\varepsilon_\mathrm{l}$ and
$\varepsilon_\mathrm{h}$ are measured to determine $\phi$, it is
important how the ATI spectrum is divided --- the effect of the second
harmonic on $P_+(E)$ and $P_-(E)$ is not equal for all $E$. To make
the selection of $\varepsilon_\mathrm{l}$ and
$\varepsilon_\mathrm{h}$, $\left[0,10\Up\right]$ was sectioned into 10
equally spaced sections, as illustrated in figure
\ref{fig:schematic-ati}. The energy was cut at $10\Up$, because it is the
highest energy the electrons can classically obtain
\citep{Paulus1995PRA}, as illustrated in figure \ref{fig:classical-paths-final-energy}(b).

For every pulse, both $\varepsilon_\mathrm{l}$ and
$\varepsilon_\mathrm{h}$ were generated from one or multiple
neighbouring sections. The sections were chosen so that
\begin{equation}
    \forall E_\mathrm{l}\in\varepsilon_\mathrm{l},E_\mathrm{h}\in\varepsilon_\mathrm{h}: E_\mathrm{l}\leq E_\mathrm{h}.
\end{equation}
A total of 495 $A_\mathrm{l}$--$A_\mathrm{h}$ representations of the
energy spectrum were generated, out of which the most useful ones were
manually selected. For more information on how the energy spectrum was
divided, see \citep{Petersson2014}.

\section{Results}

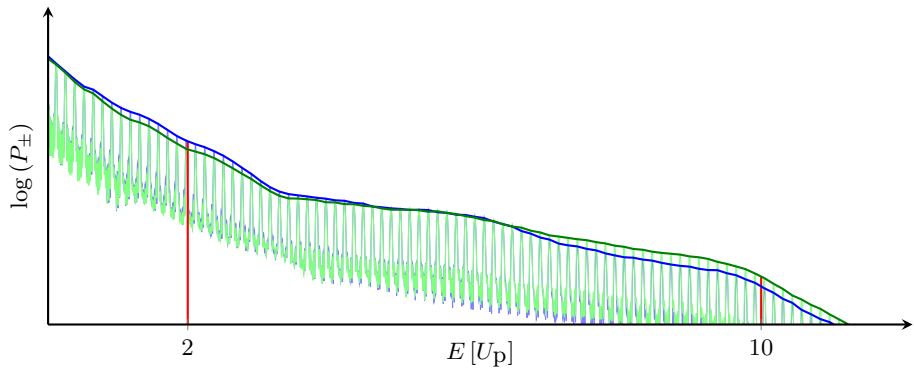
\begin{figure}[b]
  \tikzsetnextfilename{figure7}
  \begin{tikzpicture}
    \begin{axis}[
      axis line style=thick,
      ticks=none,
      width=\textwidth,
      height=\textwidth/2.25,
      xmin=0.022,
      xmax=5.32,
      axis x line=none,
      ymin=-27,
      ymax=-10,
      axis y line=none,
      ]
      \addplot [blue!50!white, thin] table[x index=0,y index=1] {resultSpectra.txt};
      \addplot [green!50!white, thin] table[x index=0,y index=2] {resultSpectra.txt};

      \addplot [red, thick] coordinates{(0.878,-36) (0.878,-17.23)};
      \addplot [red, thick] coordinates{(4.392,-36) (4.392,-24.45)};

      \addplot [blue, thick] table[x index=0,y index=1] {resultTops.txt};
      \addplot [green!50!black, thick] table[x index=0,y index=2] {resultTops.txt};
    \end{axis}
    \begin{axis}[
      axis line style=thick,
      %
      width=\textwidth,
      height=\textwidth/2.25,
      xmin=0.022,
      xmax=5.32,
      xtick={0.878,4.392},
      xticklabels={2,10},
      axis x line=bottom,
      xlabel = { \small$E\,$[$\Up$] },
      xlabel shift = -0.42cm,
      ymin=-30,
      ymax=-14,
      ytick =\empty,
      axis y line=left,
      ylabel = {$\log\left(P_\pm\right)$},
      ]
    \end{axis}
  \end{tikzpicture}
  \caption{The ATI spectra in the positive and negative directions of a two-colour field, $P_+$ and $P_-$, shown In blue and green respectively. The peaks of the spectra are outlined in darker colours. The relative intensity of the pulse used to generate the spectrum was $I_\mathrm{rel}=0.15$ and the phase difference was $\phi=\frac{\pi}{4}$.}
  \label{fig:asymmetric-ati-results}
\end{figure}

In figure \ref{fig:asymmetric-ati-results} $P_+$ and $P_-$ for a two-colour pulse are
shown. There is an asymmetry between the directions of the field,
which can be seen by observing the peaks of the spectra. For almost
high energies, $P_+$ is dominant, whereas $P_-$ dominates for low
energies. The asymmetry of electrons with an energy of $10\Up$ is
largest when the peak of the second harmonic occurs slightly after $\tau$.

\begin{figure}[b]
  \center
  \tikzsetnextfilename{figure8a_8b}
  \begin{tikzpicture}
    \tikzstyle{every node}=[font=\small]
    \begin{axis}[
      grid=both,
      major grid style={dotted, black},
      at={ (0,0) },
      width=\textwidth/3,
      height=\textwidth/3,
      xmin=-0.3,
      xmax=0.3,
      major x tick style = transparent,
      xlabel = {$A_\mathrm{l}$},
      ymin=-0.3,
      ymax=0.3,
      major y tick style = transparent,
      ylabel = {$A_\mathrm{h}$},
      y label style={rotate=270},
      ylabel shift = -0.5cm,
      ]
      \addplot [black, thick, mark=*, mark size = 1.5] table[x index=0,y index=1] {result1.txt};
      \addplot [blue!50!white, only marks, mark size = 1.0] table[x index=0,y index=1] {result1.txt};
      \node at (axis cs:-0.25,0.25) [anchor=center] {\small(a)};
    \end{axis}
    \begin{axis}[
      grid=both,
      major grid style={dotted, black},
      at={ (\textwidth/3.2,0) },
      width=\textwidth/3,
      height=\textwidth/3,
      xmin=0,
      xmax=2*pi,
      xtick={0, 3.2416, 6.2832},
      xticklabels={0, $\pi$, $2\pi$},
      xlabel = {$\theta$},
      ymin=0,
      ymax=2*pi,
      ytick={0, 3.2416, 6.2832},
      yticklabels={0, $\pi$, $2\pi$},
      ylabel = {$\phi$},
      y label style={rotate=90},
      ylabel shift = -0.1cm,
      ]
      \addplot [black, thick, mark=*, mark size = 1.5] table[x index=3,y index=2] {result1.txt};
      \addplot [blue!50!white, only marks, mark size = 1.0] table[x index=3,y index=2] {result1.txt};
      \node at (axis cs:0.55,5.78) [anchor=center] {\small(b)};
    \end{axis}
  \end{tikzpicture}
  \newline
  \tikzsetnextfilename{figure8c}
  \begin{tikzpicture}[
    declare function={
      func(\x)=(\x<=1) * (4-\x) + 
      and(\x>1, \x<=3) * (1/4*(\x - 3)^2 + 2) + 
      and(\x>3, \x<=9) * 2 +                  
      and(\x>9,\x<=10) * (2 - 1/4*(\x - 9)^2) + 
      (\x>10) * (13.5-\x)/2;
    }]
    \node at (0,2) {(c)};
    \begin{axis}[
      samples=200,
      axis line style=thick,
      height=\textwidth/4,
      width=\textwidth/1.5,
      hide x axis,
      axis x line=bottom,
      xmin=0,
      xmax=11,
      xtick=\empty,
      %
      hide y axis,
      axis y line=left,
      ymin=0,
      ymax=5,
      ytick=\empty,
      ]
      \path[name path=axis] (axis cs:0,0) -- (axis cs:11,0);
      \addplot[white,name path=f,thick,domain=0:11] {func(x)};
      \addplot[color=white!80!red] fill between[of=f and axis,soft clip={domain=0:1}];
      \addplot[color=white!80!red] fill between[of=f and axis,soft clip={domain=7:8}];
      \foreach \E/\P in {1/3, 7/2, 8/2}{
        \addplot[blue,thick] coordinates {(\E,0) (\E,\P)};
      };
      \addplot[red,thick,domain=0:10.5] {func(x)};
      \node at (axis cs:0.5,1) [anchor=center] {\small$\varepsilon_\mathrm{l}$};
      \node at (axis cs:7.5,1) [anchor=center] {\small$\varepsilon_\mathrm{h}$};
    \end{axis}
    \begin{axis}[
      samples=200,
      axis line style=thick,
      height=\textwidth/4,
      width=\textwidth/1.5,
      axis x line=bottom,
      xmin=0,
      xmax=11,
      xtick={0,1,7,8},
      xtick style={draw=none},
      xlabel={\small$E\,$[$\Up$]},
      xlabel style={at=(current axis.right of origin), anchor=north},
      axis y line=left,
      ymin=0,
      ymax=5,
      ytick=\empty,
      ylabel=${\small\log(P)}$,
      ylabel shift = -0.1cm,
      ]
    \end{axis}
  \end{tikzpicture}
  \caption{The $A_\mathrm{l}$--$A_\mathrm{h}$ representation of a
    two-colour pulse is shown in (a). In (b), the corresponding
    $\theta$--$\phi$ mapping is shown. The relative intensity was
    $I_\mathrm{rel} = 0.01$, and the energy ranges,
    $\varepsilon_\mathrm{l}=\left[0\Up,\Up\right]$ and
    $\varepsilon_\mathrm{h}=\left[7\Up,8\Up\right]$, are shown in (c).}
  \label{fig:al-ah-plane-result}
\end{figure}
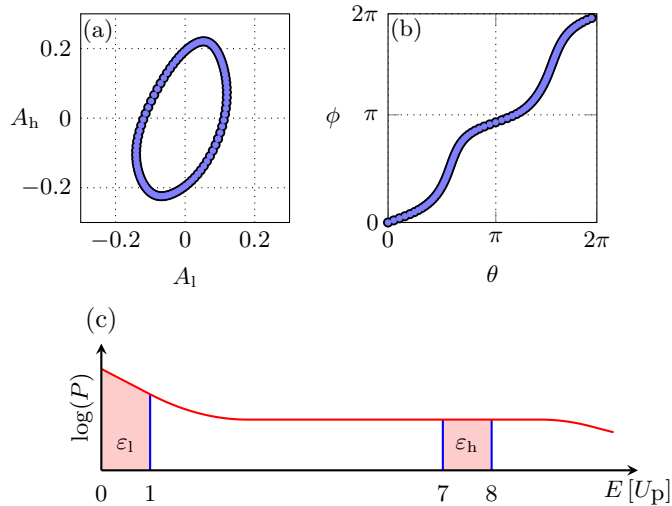

Figure \ref{fig:al-ah-plane-result}(a) shows the $A_\mathrm{l}$--$A_\mathrm{h}$
representation of a two-colour pulse. In figure \ref{fig:al-ah-plane-result}(b),
the $\theta$--$\phi$ mapping can be seen, where $\theta$ is defined as
in figure \ref{fig:al-ah-plane-schematic}(c). For the pulse and energy ranges
selected in figure \ref{fig:al-ah-plane-result},
\begin{equation}
  \left.\frac{\mathrm{d}\phi}{\mathrm{d}\theta}\right|_{\theta\in\{0,\pi\}} \ll \left.\frac{\mathrm{d}\phi}{\mathrm{d}\theta}\right|_{\theta\in\{\frac{\pi}{2},\frac{3\pi}{2}\}}.
\end{equation}

\begin{figure}[t]
  \center
  \tikzsetnextfilename{figure9a_9c}
  \begin{tikzpicture}
    \foreach \c/\file/\subfig in {0/result3a.txt/a, 1/result3b.txt/b, 2/result3c.txt/c}{
      \tikzstyle{every node}=[font=\small]
      \begin{axis}[
        grid=both,
        major grid style={dotted, black},
        at={ (\c*\textwidth/3.2,0) },
        width=\textwidth/3,
        height=\textwidth/3,
        xmin=-0.3,
        xmax=0.3,
        major x tick style = transparent,
        xlabel = {$A_\mathrm{l}$}, %
        ymin=-0.3,
        ymax=0.3,
        major y tick style = transparent,
        ylabel = {$A_\mathrm{h}$},
        y label style={rotate=270},
        ylabel shift = -0.5cm,
        ]
        \addplot [black, thick, mark=*, mark size = 1.5] table[x index=0,y index=1] {\file};
        \addplot [blue!50!white, only marks, mark size = 1.0] table[x index=0,y index=1] {\file};
        \node at (axis cs:-0.25,0.25) [anchor=center] {\small(\subfig)};
      \end{axis}
    }
  \end{tikzpicture}
  \newline
  \tikzsetnextfilename{figure9d}
  \begin{tikzpicture}[
    declare function={
      func(\x)=(\x<=1) * (4-\x) + 
      and(\x>1, \x<=3) * (1/4*(\x - 3)^2 + 2) + 
      and(\x>3, \x<=9) * 2 +                  
      and(\x>9,\x<=10) * (2 - 1/4*(\x - 9)^2) + 
      (\x>10) * (13.5-\x)/2;
    }]
    \node at (0,2) {(d)};
    \begin{axis}[
      samples=200,
      axis line style=thick,
      height=\textwidth/4,
      width=\textwidth/1.5,
      hide x axis,
      axis x line=bottom,
      xmin=0,
      xmax=11,
      xtick=\empty,
      %
      hide y axis,
      axis y line=left,
      ymin=0,
      ymax=5,
      ytick=\empty,
      ]
      \path[name path=axis] (axis cs:0,0) -- (axis cs:11,0);
      \addplot[white,name path=f,thick,domain=0:11] {func(x)};
      \addplot[color=white!80!red] fill between[of=f and axis,soft clip={domain=2:3}];
      \addplot[color=white!80!red] fill between[of=f and axis,soft clip={domain=6:7}];
      \foreach \E/\P in {2/2.25, 3/2, 6/2, 7/2}{
        \addplot[blue,thick] coordinates {(\E,0) (\E,\P)};
      };
      \addplot[red,thick,domain=0:10.5] {func(x)};
      \node at (axis cs:2.5,1) [anchor=center] {\small$\varepsilon_\mathrm{l}$};
      \node at (axis cs:6.5,1) [anchor=center] {\small$\varepsilon_\mathrm{h}$};
    \end{axis}
    \begin{axis}[
      samples=200,
      axis line style=thick,
      height=\textwidth/4,
      width=\textwidth/1.5,
      axis x line=bottom,
      xmin=0,
      xmax=11,
      xtick={2,3,6,7},
      xtick style={draw=none},
      xlabel={\small$E\,$[$\Up$]},
      xlabel style={at=(current axis.right of origin), anchor=north},
      %
      axis y line=left,
      ymin=0,
      ymax=5,
      ytick=\empty,
      ylabel=${\small\log(P)}$,
      ylabel shift = -0.1cm,
      ]
    \end{axis}
  \end{tikzpicture}
  \caption{The $A_\mathrm{l}$--$A_\mathrm{h}$ representations of three two-colour pulses. Between each figure, $I_\mathrm{rel}\in\left\{0.001,0.0125,0.15\right\}$ is increased by a factor $\approx 12.5$. In all three figures the energy ranges were $\varepsilon_\mathrm{l}=\left[2\Up,3\Up\right]$ and $\varepsilon_\mathrm{h}=\left[6\Up,7\Up\right]$, as illustrated below figures (a)--(c).}
  \label{fig:al-ah-plane-intensity}
\end{figure}
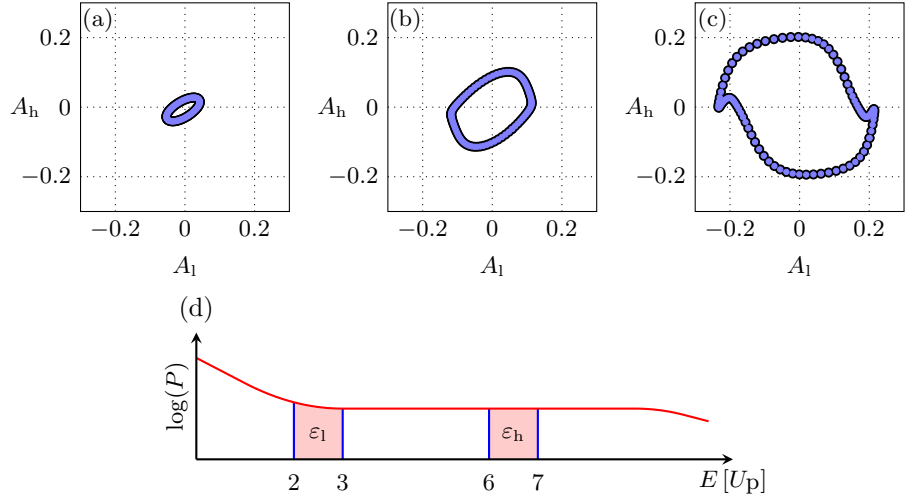

A problem which might arise due to careless selection of the
energy ranges is that $\theta\left(\phi\right)$ stops being
bijective. This can be seen in figure \ref{fig:al-ah-plane-intensity}(c), where
neither $\phi\left(0\right)$ nor $\phi\left(\pi\right)$ are unique.
This can always be circumvented by proper selection of energy ranges.

As illustrated in figure \ref{fig:al-ah-plane-intensity}, where $I_\mathrm{rel}$ is
increased exponentially between figures (a)--(c), the asymmetry
increases with $I_\mathrm{rel}$. This is to be expected, as the
addition of the second harmonic is the cause of the asymmetry, and the
radius of the $A_\mathrm{l}$--$A_\mathrm{h}$ representation can be
used to give information about the relative intensity. Note that the
$A_\mathrm{l}$--$A_\mathrm{h}$ representation can change shape as the relative
intensity increases. In figure \ref{fig:al-ah-plane-intensity}(c), the
$\theta\left(\phi\right)$ has lost the bijectivity it had for the
cases shown in figures \ref{fig:al-ah-plane-intensity}(a)--(b).

\begin{figure}[H]
  \center
  \tikzsetnextfilename{figure10a_10b}
  \begin{tikzpicture}
    \foreach \c/\file/\subfig in {0/result4a.txt/a, 1/result4b.txt/b}{
      \tikzstyle{every node}=[font=\small]
      \begin{axis}[
        grid=both,
        major grid style={dotted, black},
        at={ (\c*\textwidth/3.2,0) },
        width=\textwidth/3,
        height=\textwidth/3,
        xmin=-0.3,
        xmax=0.3,
        major x tick style = transparent,
        xlabel = {$A_\mathrm{l}$}, 
        ymin=-0.3,
        ymax=0.3,
        major y tick style = transparent,
        ylabel = {$A_\mathrm{h}$},
        y label style={rotate=270},
        ylabel shift = -0.5cm,
        ]
        \addplot [black, thick, mark=*, mark size = 1.5] table[x index=0,y index=1] {\file};
        \addplot [blue!50!white, only marks, mark size = 1.0] table[x index=0,y index=1] {\file};
        \node at (axis cs:-0.25,0.25) [anchor=center] {\small(\subfig)};
      \end{axis}
    }
  \end{tikzpicture}
  \newline
  \tikzsetnextfilename{figure10c}
  \begin{tikzpicture}[
    declare function={
      func(\x)=(\x<=1) * (4-\x) + 
      and(\x>1, \x<=3) * (1/4*(\x - 3)^2 + 2) + 
      and(\x>3, \x<=9) * 2 +                  
      and(\x>9,\x<=10) * (2 - 1/4*(\x - 9)^2) + 
      (\x>10) * (13.5-\x)/2;
    }]
    \node at (0,2) {(c)};
    \begin{axis}[
      samples=200,
      axis line style=thick,
      height=\textwidth/4,
      width=\textwidth/1.5,
      hide x axis,
      axis x line=bottom,
      xmin=0,
      xmax=11,
      xtick=\empty,
      %
      hide y axis,
      axis y line=left,
      ymin=0,
      ymax=5,
      ytick=\empty,
      ]
      \path[name path=axis] (axis cs:0,0) -- (axis cs:11,0);
      \addplot[white,name path=f,thick,domain=0:11] {func(x)};
      \addplot[color=white!80!red] fill between[of=f and axis,soft clip={domain=3:5}];
      \addplot[color=white!80!red] fill between[of=f and axis,soft clip={domain=6:7}];
      \foreach \E/\P in {3/2, 5/2, 6/2, 7/2}{
        \addplot[blue,thick] coordinates {(\E,0) (\E,\P)};
      };
      \addplot[red,thick,domain=0:10.5] {func(x)};
      \node at (axis cs:4,1) [anchor=center] {\small$\varepsilon_\mathrm{l}$};
      \node at (axis cs:6.5,1) [anchor=center] {\small$\varepsilon_\mathrm{h}$};
    \end{axis}
    \begin{axis}[
      samples=200,
      axis line style=thick,
      height=\textwidth/4,
      width=\textwidth/1.5,
      axis x line=bottom,
      xmin=0,
      xmax=11,
      xtick={3,5,6,7},
      xtick style={draw=none},
      xlabel={\small$E\,$[$\Up$]},
      xlabel style={at=(current axis.right of origin), anchor=north},
      axis y line=left,
      ymin=0,
      ymax=5,
      ytick=\empty,
      ylabel=${\small\log(P)}$,
      ylabel shift = -0.1cm,
      ]
    \end{axis}
  \end{tikzpicture}
  \caption{The $A_\mathrm{l}$--$A_\mathrm{h}$ representations of two
    short pulses, created by changing the CEP by incements. The
    pulse used to generate (a) has a Full Width at Half Maximum
    (FWHM) of 2 field cycles, whereas the one used to generate (b)
    has a FWHM of 4 field cycles. The different datapoints displayed
    in figures (a) and (b) were generated by changing the CEP
    incrementally. The bottom picture shows the studied energy
    ranges, $\varepsilon_\mathrm{l}=\left[3\Up,5\Up\right]$ and
    $\varepsilon_\mathrm{l}=\left[6\Up,7\Up\right]$. }
  \label{fig:al-ah-plane-short-pulses}
\end{figure}
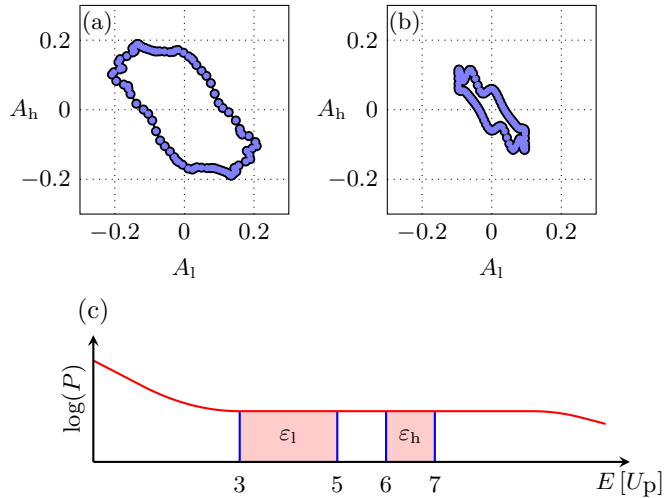

The asymmetry caused by the second harmonic in the two-colour case can
be compared to that caused by rapid change of amplitude during
few-cycle pulses --- changing the CEP affects the asymmetry of the
short pulse just as changing $\phi$ affects the asymmetry of the
two-colour field. The similarities of two-colour fields to short
pulses can be seen in figure \ref{fig:al-ah-plane-short-pulses}, where the
$A_\mathrm{l}$--$A_\mathrm{h}$ representation of two short pulses is
shown. The pulse shown in figure \ref{fig:al-ah-plane-short-pulses}(a) is of half the
duration of the one in figure \ref{fig:al-ah-plane-short-pulses}(b). It also has
significantly higher asymmetry. For short pulses the rapid amplitude
change is the cause of the asymmetry. As the amplitude gradient is
greater for short pulses, the asymmetry is as well.

\section{Conclusions}

By adding the second harmonic to a strong field, it is possible to
control the effects it has on matter. To control the effects to an as
accurate degree as possible, it is important to know the phase
difference, $\phi$,
between the two harmonics with good precision. We have shown that it
is possible to measure $\phi$
by using Stereo-ATI. The process consists of selecting two ranges of
the ATI spectrum and mapping their respective asymmetry to different
relative phases. As the asymmetry is caused by the second harmonic, it
is also possible to determine the relative intensity of the pulses by
measuring the magnitude of the asymmetry.


The second harmonic of a two-colour field can be compared to the rapid
change of amplitude of a short pulse. In both cases, the asymmetry of
each cycle results in an asymmetrical distribution of ionized
electrons. The effect is strengthened by increasing the intensity of
the second harmonic and decreasing the pulse width respectively. The
asymmetry of the electron distribution can be used to measure $\phi$
and the CEP respectively.

\ack The calculations were performed on resources provided by the
Swedish National Infrastructure for Computing (SNIC) at Lunarc, Lund
University, in project no. SNIC 2014/1-276. This work as supported in
part by funding from the NSF under grant PHY-1307083.

\bibliographystyle{iopart-num}
\bibliography{Petersson2015}

\end{document}

%% file: tikzcfg.tex
\usepackage{tikzexternal}